\documentclass[a4paper,10pt]{article}

\usepackage[naustrian,english]{babel}
\usepackage{lmodern}
\usepackage[latin1]{inputenc}
\usepackage[T1]{fontenc}
\usepackage{url}
\usepackage{algorithm}
\usepackage{amsthm}
\usepackage{amsmath}
\usepackage{amsfonts}
\usepackage{hyperref}
\usepackage{geometry}
\usepackage{tikz}
\usetikzlibrary{positioning}
\usetikzlibrary{calc}
\usetikzlibrary{arrows}

\definecolor{rectangleCol}{RGB}{215,240,255}
\definecolor{exteriorCol}{RGB}{190,220,240}
\definecolor{rose}{RGB}{250,190,190}
\definecolor{pointCol}{rgb}{0,0,1}

\tikzset{base/.style={
rectangle,
very thick,
fill=rectangleCol
}}
\tikzset{arrow/.style={
very thick,
draw=rectangleCol!70!black
}}
\tikzset{darrow/.style={
thick,
dashed,
draw=rectangleCol!70!black
}}

\newcommand{\Tma}{Theorema}
\newtheorem{definition}{Definition}
\newtheorem{theorem}{Theorem}
\newcommand{\thy}[1]{\texttt{#1}}

\title{Verifying Buchberger's Algorithm\\in Reduction Rings}
\author{Alexander Maletzky\thanks{This research was funded by the Austrian Science Fund (FWF): grant no. W1214-N15, project DK1}}
\date{\small Doctoral College ``Computational Mathematics'' and RISC\\Johannes Kepler University Linz, Austria\\\url{alexander.maletzky@dk-compmath.jku.at}}

\begin{document}
\maketitle

\begin{abstract}
In this paper we present the formal, computer-supported verification of a functional implementation of Buchberger's critical-pair/completion algorithm for computing Gr\"obner bases in reduction rings. We describe how the algorithm can be implemented and verified within one single software system, which in our case is the \Tma\ system.

In contrast to existing formal correctness proofs of Buchberger's algorithm in other systems, e.\,g. Coq and ACL2, our work is not confined to the classical setting of polynomial rings over fields, but considers the much more general setting of reduction rings; this, naturally, makes the algorithm more complicated and the verification more difficult.

The correctness proof is essentially based on some non-trivial results from the theory of reduction rings, which we formalized and formally proved as well. This formalization already consists of more than 800 interactively proved lemmas and theorems, making the elaboration an extensive example of higher-order theory exploration in \Tma.
\end{abstract}


\section{Introduction}
\label{sect::Introduction}

Buchberger's algorithm was first introduced in \cite{Buchberger1965} for computing Gr\"obner bases of ideals in polynomial rings over fields. Later, this setting was generalized to so-called \emph{reduction rings} \cite{Buchberger1984,Stifter1988}, which are essentially unital commutative rings, not necessarily free of zero divisors and not necessarily possessing any polynomial structure. The algorithm the present investigations are concerned with is a variant of Buchberger's original critical-pair/completion algorithm adapted to the reduction-ring setting. It should not come as a surprise that the increased generality of the underlying domain makes the algorithm slightly more complicated, compared to the case of polynomials over fields. The main differences will be explained in Section~\ref{sect::Algorithm}.

The theory of Gr\"obner bases, and in particular Buchberger's algorithm, has already undergone formal treatment of various kinds. For instance, the algorithm was proved correct, e.\,g., in Coq \cite{Thery2001} and ACL2 \cite{Medina-Bulo2010}. A formal analysis of its complexity in some special case was carried out by the author of this paper in \cite{Maletzky2014}, and last but not least, the algorithm could even be synthesized automatically from its specification in \cite{Craciun2008}. However, all of this was done only in the \emph{classical setting}\footnote{In this paper, the phrase ``classical setting'' always refers to the case of polynomials over fields.}, and not in the far more general setting of reduction rings. The \emph{computational} aspect of reduction rings, without formal proofs and verification of any kind, was considered in \cite{Buchberger2003}.

The software system used both for implementing the algorithm, in a functional-programming style involving pattern-matching and recursion, as well as verifying it, is the \Tma\ system \cite{Theorema}. \Tma\ is a mathematical assistant system for all phases of theory exploration: introducing new notions, designing/verifying/executing algorithms, and creating nicely structured documents.

The rest of the document is structured as follows: Section~\ref{sect::Algorithm} first defines the most important notions and then presents the algorithm in question. Section~\ref{sect::Verification} outlines the main ideas behind the computer-supported formal verification of the algorithm by means of interactive theorem proving in \Tma. Section~\ref{sect::Formalization} puts the work described here in a broader context, reporting on the underlying formal treatment of \emph{all} of reduction ring theory in \Tma; readers only interested in Buchberger's algorithm may skip this section. Finally, Section~\ref{sect::Conclusion} summarizes the content of this paper and hints on possible extensions and future work.

\section{The Algorithm}
\label{sect::Algorithm}

We now outline the algorithm under consideration. For this, let in the remainder of this paper $R$ be a \emph{reduction ring}, i.\,e. a unital commutative ring, additionally endowed with a partial Noetherian ordering $\preceq$ (among other things, which go beyond the frame of this paper). Before we can state the algorithm, we need to define the concepts of \emph{reduction} and \emph{Gr\"obner basis}:

\begin{definition}
Let $C\subseteq R$ and $a,b\in R$. Then $a$ reduces to $b$ modulo $C$, written $a\rightarrow_C b$, iff $b=a-m\,c$ for some $c\in C$ and $m\in R$, and in addition $b\prec a$. $\rightarrow_C^*$ and $\leftrightarrow_C^*$ denote the reflexive-transitive and the reflexive-symmetric-transitive closure of $\rightarrow_C$, respectively.

$C$ is called a \emph{Gr\"obner basis} iff $\rightarrow_C$ is confluent.
\end{definition}

Algorithm~\ref{alg::BB} is Buchberger's critical-pair/completion algorithm in reduction rings, given in a functional-programming style with pattern-matching and recursion, and following precisely its implementation in \Tma\ this paper is concerned with. Sticking to \Tma\ notation, function application is denoted by square brackets, tuples are denoted by angle brackets, the length of a tuple $T$ is denoted by $|T|$, and the $i$-th element of $T$ by $T_i$. Variables suffixed with three dots are so-called \emph{sequence variables} which may be instantiated by sequences of terms of any length (including $0$).

As can be seen, function \textsc{GB} only initiates the recursion by calling \textsc{GBAux} with suitable arguments. \textsc{GBAux}, in contrast, is the main function, defined recursively according to the three equations~(\ref{alg::BB::2}) (base case), (\ref{alg::BB::3}) and (\ref{alg::BB::4}). Please note that termination of \textsc{GBAux} is by no means obvious, since its second argument, which eventually has to become the empty tuple in order for the function to terminate, may be enlarged by function \textsc{update} in the second case of~(\ref{alg::BB::4}).

\begin{algorithm}[ht]
\caption{Buchberger's algorithm in reduction rings}
\label{alg::BB}
\begin{eqnarray}
\label{alg::BB::1}\textsc{GB}[C] & := & \textsc{GBAux}[C,\textsc{pairs}[|C|],1,1,\langle\rangle]\\
\label{alg::BB::2}\textsc{GBAux}[C,\langle\rangle,i,j,\langle\rangle] & := & C\\
\label{alg::BB::3}\textsc{GBAux}[C,\langle\langle k,l\rangle,r\ldots\rangle,i,j,\langle\rangle] & := & \textsc{GBAux}[C,\langle r\ldots\rangle,k,l,\textsc{cp}[C_k,C_l]]\\
\label{alg::BB::4}\textsc{GBAux}[C,P,i,j,\langle\langle b,\overline{b}\rangle,t\ldots\rangle] & := & \\\nonumber
& &\hskip-40mm \underset{h=\textsc{cpd}[b,\overline{b},i,j,C]}{\mathrm{let}}\left\{
\begin{array}{c c c}
\textsc{GBAux}[C,P,i,j,\langle t\ldots\rangle] & \Leftarrow & h=0\\
\textsc{GBAux}[\textsc{app}[C,h],\textsc{update}[P,|C|+1],i,j,\langle t\ldots\rangle] & \Leftarrow & h\neq 0
\end{array}
\right.
\end{eqnarray}
\end{algorithm}

The five arguments of function \textsc{GBAux} have the following meaning:
\begin{itemize}
\item The first argument, denoted by $C$, is the basis constructed so far, i.\,e. it serves as the accumulator of the tail-recursive algorithm. As such, it is a tuple of elements of $R$ that is initialized by the original input-tuple in equation (\ref{alg::BB::1}) and returned as the final result in equation (\ref{alg::BB::2}). Please note that the only place where it is modified is in the second case of (\ref{alg::BB::4}), where a new element $h$ is added to it (function \textsc{app}).
\item The second argument is a tuple of pairs of indices of the accumulator $C$. It contains precisely those indices corresponding to pairs of elements of $C$ that still have to be considered; hence, it is initialized by \emph{all} possible pairs of element-indices in (\ref{alg::BB::1}), using \textsc{pairs}, and updated whenever a new element is added to $C$ in (\ref{alg::BB::4}) using \textsc{update}.
\item The third and fourth arguments, denoted by $i$ and $j$, are the indices of the pair of elements of $C$ whose critical pairs are currently under investigation.
\item The last argument is the tuple containing the critical pairs of $C_i$ and $C_j$ that have not been considered so far. Once initialized by function \textsc{cp} in (\ref{alg::BB::3}), it is simply traversed from beginning to end without being enlarged by any new pairs.
\end{itemize}

The most important auxiliary function appearing in Algorithm~\ref{alg::BB} is function \textsc{cpd} in equation (\ref{alg::BB::4}). $\textsc{cpd}[b,\overline{b},i,j,C]$ returns a ring-element $h$, which is constructed by first finding $g,\overline{g}$ with $b\rightarrow_C^* g$, $\overline{b}\rightarrow_C^* \overline{g}$ and both $g$ and $\overline{g}$ irreducible modulo $C$, and then setting $h:=g-\overline{g}$. Note that $g$ and $\overline{g}$ are not unique in general, but their concrete choice is irrelevant for the correctness of \textsc{GBAux}.

Please note that ``let'' is a \Tma-built-in binder used for abbreviating terms in expressions. In equation~(\ref{alg::BB::4}) $\textsc{cpd}[b,\overline{b},i,j,C]$ is abbreviated by $h$.

Readers familiar with the classical setting might have noticed the following three differences between \textsc{GBAux} and the classical Buchberger algorithm: Firstly, not only pairs of \emph{distinct} elements $C_i$, $C_j$ have to be considered, but also pairs of identical constituents; this, in particular, implies that in reduction rings one-element sets are not automatically Gr\"obner bases. Secondly, one single pair $C_i$, $C_j$ may give rise to more than one critical pair; this is why \textsc{cp} returns a tuple of critical pairs. Thirdly, in function \textsc{cpd} it is not possible to reduce the difference $b-\overline{b}$ to normal form, even though this would be more efficient.


The algorithm as presented in this paper is not the most efficient one: as in the classical setting there are some improvements that could be applied. For instance, the so-called \emph{chain criterion} \cite{Buchberger1979} could be used to detect ``useless'' critical pairs, i.\,e. critical pairs for which $h$ in equality (\ref{alg::BB::4}) will certainly be $0$, without having to apply the (in general computation-intensive) function \textsc{cpd}. Although the chain criterion was introduced only for the case of polynomials over fields, it readily extends to reduction rings, too. Another possible improvement originating from the classical setting consists of immediately auto-reducing each element of the current basis $C$ modulo $h$ in the second case of equality (\ref{alg::BB::4}); however, it is not yet clear whether employing this improvement will always lead to correct results in the general case of arbitrary reduction rings -- this still requires further investigations.

\section{The Verification}
\label{sect::Verification}

Function \textsc{GB} has to satisfy four requirements for being totally correct: For a given tuple $C$ of elements of $R$:
\begin{enumerate}
\item the function must terminate,
\item $\textsc{GB}[C]$ has to be a tuple of elements of $R$,
\item the ideal (over $R$) generated by the elements of $\textsc{GB}[C]$ has to be the same as the ideal generated by the elements of $C$, and
\item $\textsc{GB}[C]$ has to be a Gr\"obner basis.
\end{enumerate}
The whole verification has been carried out in \Tma, using the proving capabilities of the system. More precisely, each of the four proof obligations, as well as a range of auxiliary lemmata (approx. 160; see Table~\ref{tab::TheorySummary}), have been proved \emph{interactively} in a GUI-dialog-based manner: for this, one first needs to initiate a proof attempt by setting up the initial ``proof situation'', composed of the formula one wants to prove (``proof goal'') and the list of assumptions one wants to use (``knowledge base''). The system then tries to perform some simple and obvious inference steps (e.\,g. proving implications by assuming their premises and proving their conclusions), until no more of these simple inferences are possible. Then, the user is asked to decide how to continue, i.\,e. \emph{which} of the more advanced (and maybe ``unsafe'') inferences to apply, \emph{how} to apply them in case there are several possibilities (e.\,g. providing suitable terms when instantiating a universally 
quantified assumption), and \emph{where} to continue 
in the proof search in case the current alternative does not look promising. This process is iterated until a proof is found or the proof search is aborted. Summarizing, it is really the human user who conducts the proof, but under extensive assistance by the system, which in particular ensures that all inference steps are really \emph{correct}. In the remainder of this paper, the term ``\Tma-generated proof'' always refers to precisely this type of proof.

In the following subsections we address the four proof obligations in more detail.

\subsection{Termination}
\label{sect::Termination}

As mentioned already in Section~\ref{sect::Algorithm}, termination is by no means obvious. In fact, if $R$ were not a reduction ring, termination could not even be proved, since one of the axioms characterizing reduction rings is needed only for guaranteeing termination of Algorithm~\ref{alg::BB}. The crucial point is that the second case of (\ref{alg::BB::4}) can occur only finitely often: this is guaranteed by requiring that in reduction rings there are no infinite sequences of sets $D_1,D_2,\ldots$ with $\mathrm{red}[D_i]\subset\mathrm{red}[D_{i+1}]$ for all $i\geq 1$, where $\mathrm{red}[D]$ denotes the set of \emph{reducible elements} modulo the set $D$. In the second case of (\ref{alg::BB::4}) it is easy to see that $\mathrm{red}[C]\subset\mathrm{red}[\textsc{app}[C,h]]$, meaning that this may happen only finitely many times.

Eventually, termination is proved by finding a Noetherian ordering on the set of all possible argument-quintuples that is shown to decrease in each recursive call of the function. In fact, this ordering is a lexicographic combination of the following two orderings:
\begin{enumerate}
\item ``$\trianglelefteq_1$'' is defined for subsets of $R$ as $A\triangleleft_1 B\,:\Leftrightarrow\,\mathrm{red}[B]\subset\mathrm{red}[A]$. This ordering is Noetherian because of the non-existence of certain infinite sequences in reduction rings, as sketched above.
\item ``$\trianglelefteq_2$'' is defined for arbitrary tuples as $S\triangleleft_2 T\,:\Leftrightarrow\,|S|<|T|$. Since the length of a tuple is a natural number, this ordering is clearly Noetherian.
\end{enumerate}
For comparing two argument-quintuples $(C_1,P_1,i_1,j_1,T_1)$ and $(C_2,P_2,i_2,j_2,T_2)$, first $C_1$ and $C_2$ are compared w.\,r.\,t. $\trianglelefteq_1$; if they are equal, $P_1$ and $P_2$ are compared w.\,r.\,t. $\trianglelefteq_2$; if they are equal as well, $T_1$ and $T_2$ are also compared w.\,r.\,t. $\trianglelefteq_2$ (the indices $i$ and $j$ do not play any role for termination and hence are ignored in the comparison). As one can easily see, the arguments of every recursive call of \textsc{GBAux} always decrease w.\,r.\,t. this lexicographic ordering, which furthermore is Noetherian because its constituents are.

Please note that the formal, \Tma-generated proofs of the remaining three obligations proceed by Noetherian (or ``well-founded'') induction on the set of argument-quintuples, based on the Noetherian ordering.

\subsection{Type and Ideal}
\label{sect::TypeIdeal}

The fact that $\textsc{GB}[C]$ is a tuple of elements of $R$ is obvious, since the accumulator $C$ is only modified by adding one new ring-element in the recursive call in the second case of (\ref{alg::BB::4}). Apart from that, it always remains unchanged. Furthermore, even the third requirement can be seen to be fulfilled rather easily: The element $h$ added to $C$ in the second case of (\ref{alg::BB::4}) is clearly an $R$-linear combination of elements of $C$, and hence is contained in the ideal generated by $C$. This further implies that the ideal does not change when adding $h$ to $C$.

\subsection{Gr\"obner Basis}
\label{sect::GB}

The most important requirement is the fourth one. It describes the essential property the output should have, namely being a Gr\"obner basis -- that is why the function is called \textsc{GB}. Gr\"obner bases play a very important role in computational ideal theory, since many non-trivial ideal-theoretic questions can be answered easily as soon as Gr\"obner bases for the ideals in question are known. Most importantly, ideal membership and ideal equality can simply be decided by reducing some elements to their unique normal forms modulo the given Gr\"obner bases.

For proving the fourth requirement we need one of the main results of reduction ring theory, containing a finite criterion for checking whether reduction modulo a given set, or tuple, is confluent. This result was proved formally in \Tma.

\begin{theorem}[Main Theorem of Reduction Ring Theory]
\label{thm::Main}
Let $C\subseteq R$. Then reduction modulo $C$ is confluent iff for all $c,\overline{c}\in C$ (not necessarily distinct) and all \emph{minimal non-trivial common reducibles} $a$ of $c$ and $\overline{c}$ there exists a critical pair $\langle b, \overline{b}\rangle$, with $a\rightarrow_{\{c\}} b$ and $a\rightarrow_{\{\overline{c}\}} \overline{b}$, that can be \emph{connected below} $a$ modulo $C$.
\end{theorem}

The statement of Theorem~\ref{thm::Main} is somewhat vague. For the precise definition of \emph{minimal non-trivial common reducible} we have to refer the interested reader to \cite{Buchberger1984,Stifter1988} or to our recent technical report \cite{Maletzky2015a}. Only note that in the classical setting the minimal non-trivial common reducible of two polynomials $p$ and $q$ is precisely the least common multiple of the leading monomials of $p$ and $q$. Finally, two elements $b,\overline{b}\in R$ are \emph{connectible below} another element $a$ modulo $C$ iff $b\leftrightarrow_C^* \overline{b}$ and every intermediate element in the chain of reductions is strictly smaller than $a$ w.\,r.\,t. $\preceq$.

Now we can outline how function \textsc{GB} can be shown to satisfy the fourth requirement: In (\ref{alg::BB::4}), if the element $h$ constructed by \textsc{cpd} is $0$ then the critical pair $\langle b,\overline{b}\rangle$ can be connected (modulo $C$) below the minimal non-trivial common reducible $a$ of $C_i$ and $C_j$ it corresponds to; otherwise, $b$ and $\overline{b}$ can certainly be connected below $a$ modulo the enlarged tuple $\textsc{app}[C,h]$ (second case). Hence, in either case the critical situation corresponding to $\langle b,\overline{b}\rangle$ is resolved, and the algorithm proceeds with the next critical pair of $C_i$ and $C_j$, unless all of them have already been dealt with; in that case, the next pair of elements of $C$ is considered (equation (\ref{alg::BB::3})). Termination guarantees that at some point \emph{all} pairs of elements of $C$ have been dealt with (even those added in (\ref{alg::BB::4})), such that in the end the criterion of Theorem~\ref{thm::Main} is fulfilled and the 
output returned by the algorithm is indeed a Gr\"obner basis.

More formally, the crucial property of \textsc{GBAux} is the following:
\begin{theorem}
\label{thm::correctness}
For all tuples $C$ of elements in $R$, all index-pair tuples $P$, all indices $i$ and $j$, and all critical-pair tuples $M$: The result $G$ of $\textsc{GBAux}[C,P,i,j,M]$ is again a tuple of elements of $R$ such that \emph{all} critical pairs of \emph{all} $C_k,C_l$, for $\langle k,l\rangle\in P$, can be connected below their corresponding minimal non-trivial common reducibles modulo $G$, and the same is true also for the critical pairs in $M$.
\end{theorem}

As mentioned at the end of Section~\ref{sect::Termination}, the interactively generated \Tma-proof of Theorem~\ref{thm::correctness} proceeds by Noetherian induction on the set of all input-quintuples, distinguishing four cases based on the shape of the input arguments, according to the left-hand-sides of the three equalities (\ref{alg::BB::2}), (\ref{alg::BB::3}) and (\ref{alg::BB::4}) (where the case corresponding to (\ref{alg::BB::4}) is split into two subcases depending on whether $h=0$ or not).

The total effort for first formalizing and then verifying Algorthm~\ref{alg::BB}, already knowing Theorem~\ref{thm::Main}, was approximately 70 working hours. As can be seen in Table~\ref{tab::TheorySummary} of the next section, the number of formulas that had to be proved for that purpose is 165.

\section{The Formal Treatment of Reduction Ring Theory}
\label{sect::Formalization}

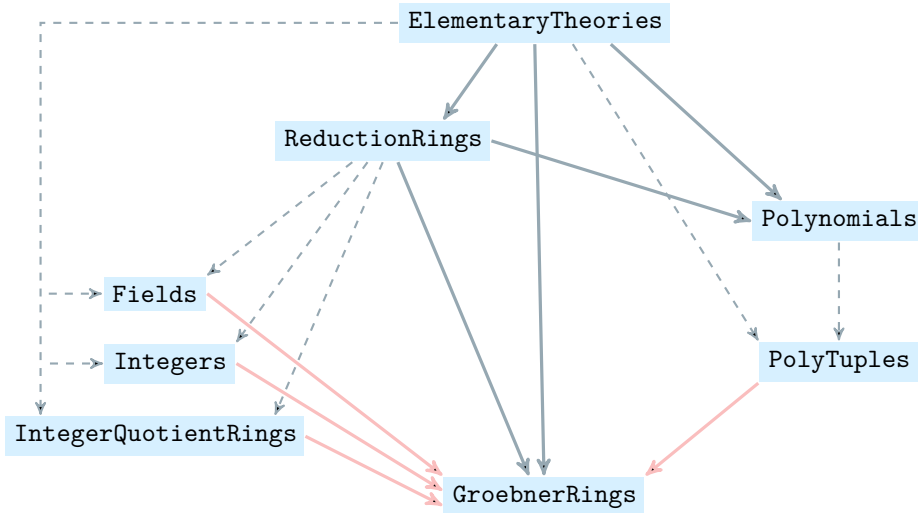
\begin{figure}
\centering
\begin{tikzpicture}[text centered, node distance=10mm and 8mm]
\node (et) [base] {\thy{ElementaryTheories}};
\node (rr) [base, below=of et, xshift=-20mm] {\thy{ReductionRings}};
\node (py) [base, below=of rr, xshift=60mm, yshift=5mm] {\thy{Polynomials}};
\node (fl) [base, below=of rr, xshift=-30mm, yshift=-5mm] {\thy{Fields}};
\node (in) [base, below=of fl, xshift=2mm, yshift=6mm] {\thy{Integers}};
\node (qr) [base, below=of in, xshift=-2mm, yshift=6mm] {\thy{IntegerQuotientRings}};
\node (pt) [base, below=of py, yshift=-3mm] {\thy{PolyTuples}};
\node (gr) [base, right=of qr, xshift=10mm, yshift=-8mm] {\thy{GroebnerRings}};
\pgfsetarrowsend{stealth'}
\draw[arrow] ($(et.south)-(5mm,0mm)$) -- ($(rr.north)+(8mm,0mm)$);
\draw[arrow] ($(et.south)+(10mm,0mm)$) -- ($(py.north west)+(4mm,0mm)$);
\draw[darrow] ($(et.south)+(5mm,0mm)$) -- (pt.north west);
\draw[darrow] (et.west) -| ($(qr.north)-(15mm,0mm)$);
\draw[darrow] ($(fl.west)-(7mm,0)$) -- (fl.west);
\draw[darrow] ($(in.west)-(7mm,0)$) -- (in.west);
\draw[arrow] (et.south) -- (gr.north);
\draw[arrow] (rr.east) -- (py.west);
\draw[darrow] ($(rr.south)-(4mm,0mm)$) -- (fl.north east);
\draw[darrow] ($(rr.south)-(2mm,0mm)$) -- (in.north east);
\draw[darrow] (rr.south) -- ($(qr.north east)-(4mm,0mm)$);
\draw[arrow] ($(rr.south)+(2mm,0mm)$) -- ($(gr.north)-(2mm,0mm)$);
\draw[darrow] (py.south) -- (pt.north);

\draw[arrow,draw=rose] (fl.east) -- (gr.north west);
\draw[arrow,draw=rose] (in.east) -- ($(gr.north west)-(0mm,2mm)$);
\draw[arrow,draw=rose] (qr.east) -- ($(gr.north west)-(0mm,4mm)$);
\draw[arrow,draw=rose] (pt.south west) -- (gr.north east);
\end{tikzpicture}
\caption{\small The structure of the formalization. An arrow from $A$ to $B$ denotes dependency of $B$ on $A$, in the sense that formulas from $A$ are used in $B$ in proofs (gray) or computations (red). Dashed arrows denote future dependencies.}
\label{fig::Theory}
\end{figure}

What has been presented in the previous sections of this paper actually only constitutes a small fragment of a much larger endeavor: The formalization and formal verification of the theory of reduction rings in \Tma. This project was started two years ago with the aim of representing all aspects of the theory, both theoretic and algorithmic, in a unified and -- most importantly -- certified way in a computer system. At the moment the whole formalization consists of eight individual components, each being a separate \Tma\ notebook containing definitions, theorems and algorithms of a particular part of reduction ring theory. Figure~\ref{fig::Theory}, which is taken from \cite{Maletzky2015a}, shows the entire theory graph with all components and their dependencies on each other. The algorithm this paper is concerned with, as well as its correctness proof as described in the previous section, is contained in theory \thy{GroebnerRings}, whereas Theorem~\ref{thm::Main} together with its proof are contained in  theory \thy{ReductionRings}. For more information on the formalization the interested reader is referred to \cite{Maletzky2015a}; however, note that there the correctness-proof of Buchberger's algorithm is still labelled as ``future work'', because the proof has been completed only \emph{after} writing the report.

As can be seen from the dashed arrows in Figure~\ref{fig::Theory}, the formal verification of some parts of the theory still awaits its completion: the proofs that certain basic domains, namely fields, the integers, integer quotient rings, and polynomials represented as tuples of monomials, are reduction rings have not been carried out yet. This is not because these proofs turned out to be extraordinarily difficult, but rather the opposite: we do not expect any major difficulties there and instead focused on the far more involved proofs (in \thy{ReductionRings}, \thy{Polynomials} and \thy{GroebnerRings}) first, just to be sure that everything works out as it is supposed to. After all, the correctness of Buchberger's algorithm is absolutely independent of theories \thy{Fields}, \thy{Integers}, etc.

Table~\ref{tab::TheorySummary} lists the sizes of the individual components of the formalization in terms of the numbers of formulas, the numbers of proofs, and the average and maximum proof sizes. Summing things up one arrives at almost 1700 formulas and more than 1100 interactively-generated proofs in the formalization, making it an extensive piece of computerized mathematics.

\begin{table}[ht]
\centering
\begin{tabular}{l r r r}
Theory & Formulas & Proofs & Proof Size (avg./max.)\\\hline
\thy{ElementaryTheories} & 630 & 390 & 21.9 / 137\\
\thy{ReductionRings} & 315 & 253 & 38.1 / 198\\
\thy{Polynomials} & 397 & 341 & 45.8 / 322\\
\thy{GroebnerRings} & 226 & 165 & 37.0 / 154\\
\thy{Fields} & 17 & 0\\
\thy{Integers} & 20 & 0\\
\thy{IntegerQuotientRings} & 19 & 0\\
\thy{PolyTuples} & 66 & 0\\\hline
& \textbf{1690} & \textbf{1149} & \textbf{34.7} / \textbf{322}
\end{tabular}
\caption{\small Number of formulas and proofs in the formalization. The proof size refers to the number of inference steps.}
\label{tab::TheorySummary}
\end{table}

\section{Conclusion}
\label{sect::Conclusion}

On the preceding pages we described the implementation and formal verification of a non-trivial algorithm of high relevance in computational ideal theory. Although the work is of interest on its own, it also serves as a major case study in how program verification, including the formal development of the underlying theories, can effectively be carried out in the \Tma\ system. In addition, most of the elementary mathematical concepts formalized for the present verification, like tuples, (lexicographic) orders and infinite sequences, can be reused for the \Tma-verification of algorithms and programs in completely different areas in the future.

The work described in this paper also revealed a potential improvement of \Tma: Correctness proofs of functional programs are typically achieved following a fixed set of steps, consisting of finding termination orders, proving specialized induction schemas, and using these schemas to prove that certain properties hold for the function, provided they hold for each recursive call. At present, these steps have to be carried out manually, but it is clearly possible to automate the process at least in \emph{some} way -- just as in the well-known Isabelle system \cite{Isabelle}.

\paragraph*{Acknowledgements}
I thank Bruno Buchberger and Wolfgang Windsteiger for many inspiring discussions about Gr\"obner bases and \Tma, and I also thank the anonymous referees for their valuable comments and suggestions.

This research was funded by the Austrian Science Fund (FWF): grant no. W1214-N15, project DK1.

\bibliography{References}
\end{document}